\documentclass[aps,prx,10pt,twocolumn,superscriptaddress]{revtex4-2}

\usepackage{times}
\usepackage{changes}

\usepackage{graphicx}
\usepackage{amssymb}
\usepackage{epstopdf}
\usepackage{amsmath}
\usepackage{color}
\usepackage{hyperref}
\usepackage{bbold}

\usepackage[acronym]{glossaries}
\usepackage{physics}
\makeatletter

\graphicspath{{Images/}}

\hypersetup{
	pdfpagemode={UseOutlines},
	bookmarksopen,
	pdfstartview={FitH},
	colorlinks,
	linkcolor={black},
	citecolor={blue},
	urlcolor={blue}}

\usepackage{hyperref}
\hypersetup{
	colorlinks=true,
	linkcolor=black,          
	citecolor=blue,           
	filecolor=magenta,        
	urlcolor=cyan
}

\makeatother

\begin{document}
	
	\title{Bistability and time crystals in long-ranged directed percolation}
	
	\author{Andrea Pizzi}
	\affiliation{Cavendish Laboratory, University of Cambridge, Cambridge CB3 0HE, United Kingdom}
	\author{Andreas Nunnenkamp}
	\affiliation{School of Physics and Astronomy and Centre for the Mathematics and Theoretical Physics of Quantum Non-Equilibrium Systems, University of Nottingham, Nottingham, NG7 2RD, United Kingdom}
	\author{Johannes Knolle}
	\email[email: ]{j.knolle@tum.de}
	\affiliation{Department of Physics, Technische Universit{\"a}t M{\"u}nchen, James-Franck-Stra{\ss}e 1, D-85748 Garching, Germany}
	\affiliation{Munich Center for Quantum Science and Technology (MCQST), 80799 Munich, Germany}
	\affiliation{Blackett Laboratory, Imperial College London, London SW7 2AZ, United Kingdom}
	
	\maketitle
	
	\section*{Abstract}
	Stochastic processes govern the time evolution of a huge variety of realistic systems throughout the sciences. A minimal description of noisy many-particle systems within a Markovian picture and with a notion of spatial dimension is given by probabilistic cellular automata, which typically feature time-independent and short-ranged update rules. Here, we propose a simple cellular automaton with power-law interactions that gives rise to a bistable phase of long-ranged directed percolation whose long-time behaviour is not only dictated by the system dynamics, but also by the initial conditions. In the presence of a periodic modulation of the update rules, we find that the system responds with a period larger than that of the modulation for an exponentially (in system size) long time. This breaking of discrete time translation symmetry of the underlying dynamics is enabled by a self-correcting mechanism of the long-ranged interactions which compensates noise-induced imperfections. Our work thus provides a firm example of a classical discrete time crystal phase of matter and paves the way for the study of novel non-equilibrium phases in the unexplored field of `Floquet probabilistic cellular automata'.
	
	\section{Introduction}
	Percolation theory describes the connectivity of networks, with applications pervading virtually any branch of science \cite{stauffer2018introduction}, including economics \cite{stauffer2001percolation}, engineering \cite{lee2011trap}, neurosciences \cite{breskin2006percolation}, social sciences \cite{morone2015influence}, geoscience \cite{king2001predicting}, food science \cite{illy2005espresso} and, most prominently, epidemiology \cite{moore2000epidemics}.
	Among the multitude of phenomena described by percolation, of predominant importance are spreading processes, in which time plays a crucial role and that can be studied within models of directed percolation (DP) \cite{hinrichsen2000non}. Characterized by universal scalings in time \cite{odor2004universality}, in their discretized versions these models are probabilistic cellular automata (PCA), that is, dynamical systems with a state evolving in discrete time according to a set of stochastic and generally short-ranged update rules. To account for certain realistic situations, e.g., of long-distance travels in epidemic spreading, DP has been extended to long-ranged updates \cite{mollison1977spatial,cannas1998phase} leading to a change of the universal scaling exponents \cite{janssen1999levy}. 
	
	Despite their wide applicability, PCAs have surprisingly remained an outlier in a branch of non-equilibrium physics that has recently experienced a tremendous amount of excitement -- that of discrete time crystals (DTCs) \cite{wilczek2012quantum, shapere2012classical, sacha2015modeling, khemani2016phase, else2016floquet, yao2017discrete, moessner2017equilibration}. In essence, DTCs are systems that, under the action of a time-periodic modulation with period $T$, exhibit a periodic response at a different period $T' \neq T$, thus breaking the discrete time-translational symmetry of the drive and of the equations of motion. DTCs thus extend the fundamental idea of symmetry breaking \cite{sachdev2007quantum} to non-equilibrium phases of matter. Following the pioneering proposals in the context of many-body-localised (MBL) systems \cite{khemani2016phase, else2016floquet}, DTCs have been observed experimentally \cite{choi2017observation, zhang2017observation}, and their notion has been extended beyond MBL \cite{gong2018discrete, gambetta2019discrete, pizzi2019higher, machado2020long}.
	
	More recently, Yao and collaborators have fleshed out the essential ingredients of a classical DTC phase of matter \cite{yao2020classical}. Namely, in a classical DTC, many-body interactions should allow for an infinite autocorrelation time, which should be stable in the presence of a noisy environment at finite temperature, a subtle requirement that rules out the vast class of long-known deterministic dynamical systems. Despite various efforts \cite{yao2020classical, Gambetta2019, khasseh2019many, heugel2019classical}, an example of such a classical DTC has mostly remained elusive, and proving an infinite autocorrelation time robust to noise and perturbations for this phase of matter is an outstanding problem. The general expectation is in fact that PCAs and other minimal models for noisy systems in one spatial dimension can only show a transient subharmonic response because noise-induced imperfections generically nucleate and spread, destroying true infinite-range symmetry breaking in time \cite{yao2020classical, buvsic2012density}.
	
	Here, we overcome these difficulties by introducing a simple and natural generalization of DP in which the dynamical rules are governed by power-law correlations. This leads to qualitative changes of the system behaviour and, crucially, the emergence of a bistable phase of long-ranged DP, enabled by the ability of long-range interactions to counteract the dynamic proliferation of defects. By adding a periodic modulation to the update rules, we then study a version of `Floquet DP' and show that the underlying bistable phase  intimately connects to a stable DTC. In this non-equilibrium phase, the system is able to self-correct noise-induced errors and the autocorrelation time grows exponentially with the system size, thus becoming infinite in the thermodynamic limit. 
	In analogy to the one-dimensional Ising model for which, at equilibrium, long-range interactions enable a normally forbidden finite-temperature magnetic phase \cite{dyson1969existence,thouless1969long}, in our model, out of equilibrium, the long-range interactions lead to a classical time-crystalline phase. Crucially, our results appear naturally in a minimal model of long-ranged DP, but are expected to find applications in many different contexts of dynamical many-body systems.
 	
 	Basic understanding of new concepts has historically been built around the study of minimal models, such as the Ising model for magnetism at equilibrium \cite{dyson1969existence,thouless1969long}, the kicked transverse field Ising chain for DTCs \cite{khemani2016phase,else2016floquet}, or the prototypical Domany-Kinzel (DK) PCA for DP \cite{domany1984equivalence}.
	In this paper, we start our discussion with a brief review of the DK model and then generalize it to include power-law interactions. We characterize its phase diagram, and show that its long-range nature is the key ingredient for the emergence of a bistable phase. Finally, we include a periodic drive for the long-ranged DP process and show with a careful scaling analysis that the autocorrelation time of the subharmonic response is exponential in system size. In the thermodynamic limit, our model provides therefore the first example of a PCA behaving as a classical DTC, which is persistent and stable to the continuous presence of noise. Lastly, we conclude with a summary of our findings and an outlook for future research.
	
	\section{Results}
\begin{figure}[t]
		\begin{center}
			\includegraphics[width=\linewidth]{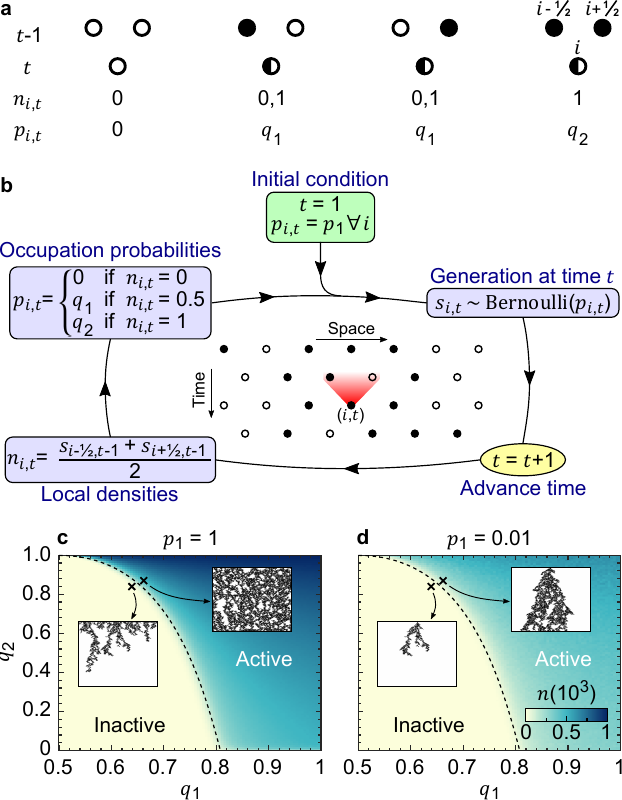}
		\end{center}
		\vskip -0.5cm \protect\caption
		{\textbf{Domany-Kinzel model of directed percolation.}
			(a) The probability $p_{i,t}$ of site $i$ to be occupied at time $t$ depends on the occupation of its nearest-neighbours $i\pm \frac{1}{2}$ at time $t-1$ and can take discrete values $0$, $q_1$, and $q_2$.
			(b) Flowchart representation of the DK model. The initial occupation probability is uniform $p_{i,t=1} = p_1$. At time $t$, each site $i$ is either occupied ($s_{i,t} = 1$) or empty ($s_{i,t} = 0$) with probability $p_{i,t}$ and $1-p_{i,t}$, respectively. Time is advanced and local densities $\{n_{i,t}\}_i$ are computed for each site $i$ as averages of the nearest-neighbour occupations at previous time, and these densities determine the occupation probabilities for the next generation, see Eq.~\eqref{eq. DK p}. The generations at all subsequent times are obtained by iteration.
			(c,d) The density $n$ at late times can be used to discern the active and inactive phases, in which $n(t = 10^3) > 0$ and $\approx 0$, respectively. The dashed lines serve as a reference to locate the phase boundary, and are the same for initial densities $p_1 = 1$ (c) and $p_1 = 0.01$ (d). The insets show representative single instances of the DP for the points in the $(q_1, q_2)$ plane marked with a cross. Here, $L = 100$ and $R = 10^3$.}
		\label{fig1}
	\end{figure}

	\textbf{Review of directed percolation.}
	We consider a triangular lattice in which one dimension can be interpreted as discrete space $i$ and the other one as discrete time $t = 1,2,3,\dots$, see Fig.~\ref{fig1}. To implicitly account for the triangular nature of the lattice, $i$ runs over integers and half-integers at odd and even times $t$, respectively. We denote $L$ the spatial system size, and are interested in the thermodynamic limit $L \to \infty$. The site $i$ at time $t$ can be either occupied or empty, 
	$s_{i,t} = 0,1$.
	For a given time $t$, we call \emph{generation} the collection of variables $\{s_{i,t}\}_i$ specifying the system state.
	Initially, the sites are occupied with uniform probability $p_1 > 0$. A DP process is defined by a stochastic Markovian update rule with which, starting from the initial generation $\{s_{i,1}\}_i$, all subsequent generations $\{s_{i,t}\}_i$ are obtained one by one. The main observable we will focus on is the global density $n(t)$ (henceforth just referred to as \emph{density} for brevity) defined as
	\begin{equation}
	n(t) = \langle \langle s_{i,t} \rangle_{i} \rangle_{\text{runs}}
	\label{eq. n}
	\end{equation}
	where the inner and outer brackets denote average over the $L$ sites and over $R$ independent runs, respectively. Since $n(1) = p_1$, we will often refer to $p_1$ as initial density.
	
	The simplest, and yet already remarkably rich, example of the above setting of DP is the DK model \cite{domany1984equivalence}. Here, we briefly review it adopting an unconventional notation that, making explicit use of a local density, will prove very convenient for a straightforward generalization to a model of long-ranged DP. 
	
	In the DK model, the probability of site $i$ to be occupied at time $t$ depends on the state of its neighbours $i\pm1/2$ at previous time $t-1$. More specifically, as summarized in Fig.~\ref{fig1}(a), site $i$ is: (i) empty if both its neighbours were empty, (ii) occupied with probability $q_1$ if one and just one of its neighbours was occupied, (iii) occupied with probability $q_2$ if both its neighbours were occupied. To account for these possibilities in a compact fashion, we define a \emph{local density} $n_{i,t}$ as
	\begin{equation}
	n_{i,t} = \frac{s_{i-\frac{1}{2}, t-1} + s_{i+\frac{1}{2}, t-1}}{2},
	\label{eq. loc density DK}
	\end{equation}
	and say that site $i$ at time $t$ is occupied with a probability $p_{i,t}$ given by
	\begin{equation}
	p_{i,t} =\begin{cases}
	0   &\text{if} \quad n_{i,t} =  0\\
	q_1 &\text{if} \quad n_{i,t} =  0.5 \\
	q_2 &\text{if} \quad n_{i,t} =  1.
	\end{cases}
	\label{eq. DK p}
	\end{equation}
	In other words, the probability $p_{i,t}$ is a nonlinear function $f_{q_1, q_2}(n_{i,t})$ of the local density $n_{i,t}$, with domain $\{0, 0.5, 1\}$. Since $n_{i,t}$ only involves the nearest neighbours of site $i$, the DK model of DP is obviously `short-ranged'. In essence, $s_{i,t}$ is a Bernoullian random variable of parameter $p_{i,t}$, which we compactly denote $s_{i,t} \sim \text{Bernoulli}\left(p_{i,t}\right)$. The complexity of this model arises from the fact that the value of the parameter $p_{i,t}$ is not known \textit{a priori}, as it depends on the actual state of the system at previous time $t-1$. Equipped with a random number generator, one can obtain all the generations one by one according to the above procedure, as schematically illustrated in the flowchart of Fig.~\ref{fig1}(b). Reiterating for several independent runs, one finally obtains the time series of the density $n$ in Eq.~\eqref{eq. n}.	
	
	The DK model features two dynamical phases, shown in Fig.~\ref{fig1}(c,d). In the {\it inactive phase}, for small enough probabilities $q_1$ and $q_2$, the system eventually reaches the completely unoccupied absorbing state, that is no percolation occurs. In the {\it active phase} instead, for large enough probabilities $q_1$ and $q_2$, a finite fraction of sites remains occupied up to infinite time, that is the system percolates. For small initial probability $p_1 \ll 1$, the critical line separating the two phases is characterized by a power-law growth of the density \cite{Odor2004}, $n\sim t^{\theta}$, with exponent $\theta \approx 0.31$. As conjectured by Grassberger ~\cite{grassberger1995damage}, this exponent is universal for all systems in the DP universality class. Indeed, DP exemplifies how the unifying concept of universality pertaining to quantum and classical many-body systems \cite{Stanley1999} can be extended to non-equilibrium phenomena.
	
	Important for our work is that, in the DK model, whether the system percolates or not depends on the parameters $q_1$ and $q_2$, but not on the initial density $p_1$, at least as long as $p_1>0$. Indeed, the phase boundaries for initial densities $p_1 = 0.01$ and $p_1 = 1$ in Fig.~\ref{fig1}(c) and Fig.~\ref{fig1}(d), respectively, coincide.
	
	\begin{figure*}[t]
		\begin{center}
			\includegraphics[width=\linewidth]{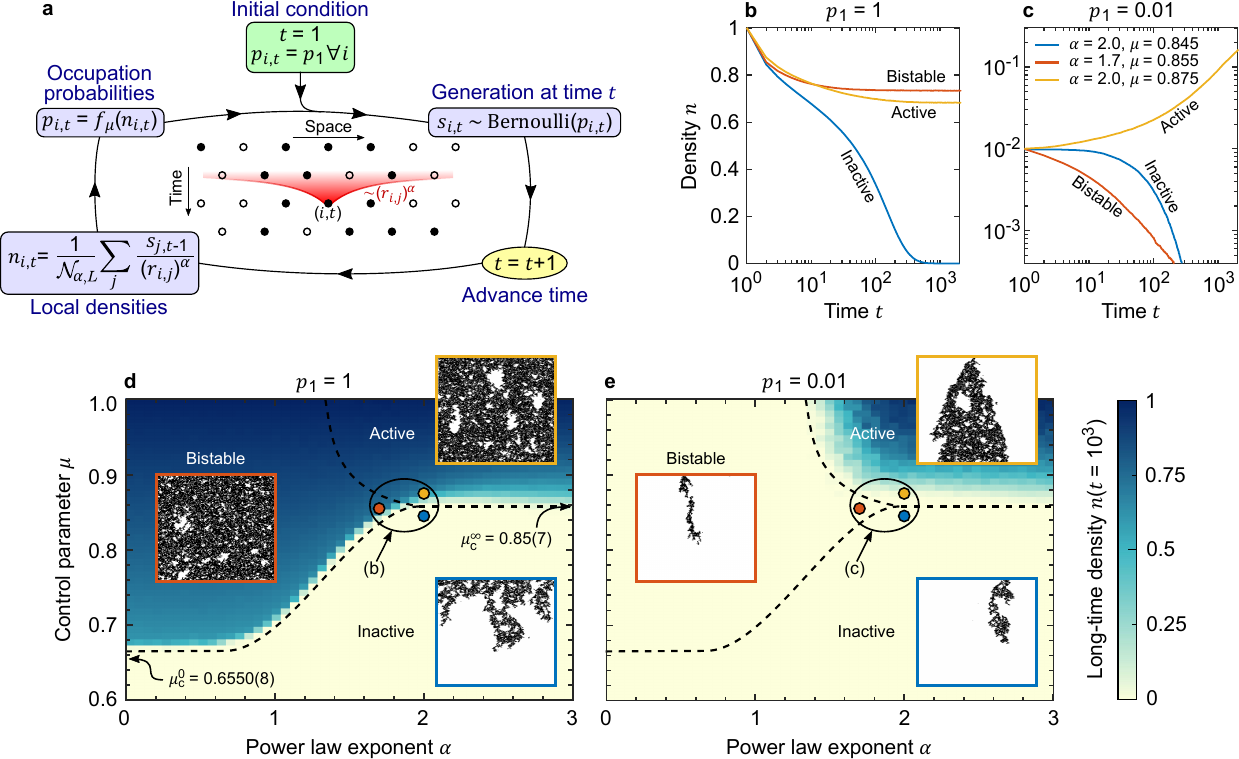}\\
		\end{center}
		\vskip -0.5cm \protect
		\caption{
			\textbf{Long-ranged directed percolation and bistability.} (a) Flowchart representation of the long-ranged DP. Starting from an initial condition $p_{i,t=1} = p_1$, site $i$ at time $t$ is occupied ($s_{i,t} = 1$) with probability $p_{i,t}$. Local densities $\{n_{i,t}\}_i$ are computed as power-law-weighted averages of the previous generation $\{s_{i,t-1}\}_i$, and the occupation probabilities are updated as $p_{i,t} = f_\mu\left(n_{i,t} \right)$, see Eqs.~\eqref{eq. loc density} and \eqref{eq. f}. (b,c) Time evolution of the density $n$ for $p_1 = 1$ (b) and $p_1 = 0.01$ (c) for various representative values of the power-law exponent $\alpha$ and control parameter $\mu$. Three dynamical phases can be distinguished: (i) inactive -- the density $n$ decays to $0$ (blue); (ii) active -- $n$ does not decay to $0$ (yellow); (iii) bistable -- $n$ either decays to $0$ or not depending on whether the initial density $p_1$ is small or large (red). (d,e) Long-time density $n(t = 10^3)$ in the plane of $\alpha$ and $\mu$ for $p_1 = 1$ (d) and $p_1 = 0.01$ (e). With the criterion used in (b,c), we discern the three phases: inactive (light), active (dark), and bistable (light or dark depending on $p_1$). The dashed lines help locating the phases and coincide in (d) and (e), and critical values $\mu_c^0$ and $\mu_c^\infty$ of $\mu$ in the limits $\alpha \to 0$ and $\alpha \to \infty$, respectively, are reported (the offset of $\mu_c^0$ from the dashed line, as well as the softening of the dashed line for $\alpha \approx 1$, are due to finite-size effects). Crucially, the bistable phase is present only for small enough $\alpha \lessapprox 2$, that is for a sufficiently long-ranged DP. Single instances of the DP for the three phases are shown in the insets, as obtained for the $\alpha$ and $\mu$ indicated with colored dots, and corresponding to the parameters used in (b,c). Here, $R = 10^4$ and $10^2$ in (b,c) and (d,e), respectively, and $L = 500$.}
		\label{fig2}
	\end{figure*}
	
	\textbf{Long-ranged percolation and bistability.}
	As the vast majority of PCA, the DK model features short-ranged update rules \cite{hinrichsen2000non}. In realistic systems, however, it is often the case that the occupation of a site $i$ is influenced not only by the neighbouring sites, but also by farther sites $j$, with an effect decreasing with the distance $r_{i,j}$ between the sites. Building on an analogy with the DK model, we propose here a model for such a `long-ranged' DP, whose protocol is explained in the flowchart of Fig.~\ref{fig2}(a). Specifically, we consider as a local density $n_{i,t}$ a power-law-weighted average of the previous generation $\{s_{j,t-1}\}_j$ centered around site $i$
	\begin{equation}
	n_{i,t} = \frac{1}{\mathcal{N}_{\alpha, L}} \sum_{j} \frac{s_{j,t-1}}{\left(r_{i,j}\right)^\alpha},
	\label{eq. loc density}
	\end{equation}
	where the normalization factor $\mathcal{N}_{\alpha, L}$ ensures $n_{i,t} = 1$ if all sites $j$ are occupied, and the adjective `local' emphasizes the site-dependence. The occupation probability $p_{i,t}$ then depends on the local density $n_{i,t}$ through some nonlinear function $f_\mu$, that for concreteness we consider to be
	\begin{equation}
	p_{i,t} = \mu \tanh(4n_{i,t}^2),
	\label{eq. f}
	\end{equation}
	with $\mu \in (0,1)$ a control parameter. The whole DP dynamics is determined via the occupations $s_{i,t} \sim \text{Bernoulli}\left(p_{i,t}\right)$ and reiterating from one generation to the next. Note, our findings are not contingent on the specific choice of Eqs.~\eqref{eq. loc density} and \eqref{eq. f}, but are rather expected to hold generally for a broad class of long-ranged forms of the densities $n_{i,t}$ and of functions $f_\mu$ -- see Section Methods for details.
	
	We emphasise that Eq.~\eqref{eq. loc density}, Eq.~\eqref{eq. f}, and the flowchart in Fig.~\ref{fig2}(a) are a natural generalization of Eq.~\eqref{eq. loc density DK}, Eq.~\eqref{eq. DK p}, and Fig.~\ref{fig1}(b), respectively. Furthermore, whereas in the DK model the control parameters are the probabilities $q_1$ and $q_2$, the control parameter is now $\mu$. As an important difference, now the domain of $f_\mu$ accounts for several (and $\alpha$-dependent) values of $n_{i,t}$, for which the piecewise definition of $p_{i,t}$ as in Eq.~\eqref{eq. DK p} would have been unpractical, and the compact form of Eq.~\eqref{eq. f} was necessary instead.
	
	The introduction of a long-ranged local density $n_{i,t}$ in Eq.~\eqref{eq. loc density} has profound implications. Arguably, the most dramatic is the appearance of a bistable phase, in addition to  the standard active and inactive ones. In the bistable phase, the ability of the system to percolate depends on the initial density $p_1$, see the red lines in Fig.~\ref{fig2}(b,c). That is, the bistable phase features two basins of attraction, resulting into an asymptotically vanishing or finite $n$, respectively, and separated by some critical initial density $p_{1,c} > 0$. To characterize systematically the dynamical phases of our model, we plot in Fig.~\ref{fig2}(d,e) the long-time density $n(t=10^3)$ as a suitable order parameter in the plane of the power-law exponent $\alpha$ and control parameter $\mu$. Comparing the results obtained for a large and a small initial density $p_1$, it is possible to sketch a phase diagram composed of three phases: (i) inactive -- $n$ decays to $0$ at long times; (ii) active -- $n$ does not decay at long times; (iii) bistable -- $n$ either decays or not depending on $p_1$ being small or large. The existence of this bistable phase is in striking contrast with short-ranged models of DP such as the DK model, and in fact appears only for $\alpha \lessapprox 2$, that is, when the local densities $\{n_{i,t}\}_i$ are correlated over a sufficiently long range. To understand the origin of this rich phenomenology, we study the short- and infinite-range limits of our DP process.

	In the short-range limit $\alpha \to \infty$, the local densities $n_{i,t}$ reduce to the averages of the nearest-neighbour occupations $s_{i-\frac{1}{2},t-1}$ and $s_{i+\frac{1}{2},t-1}$, that is, Eq.~\eqref{eq. loc density} recasts into Eq.~\eqref{eq. loc density DK} and the DK model is recovered. In the notation of Eq.~\eqref{eq. DK p}, the DK parameters are $q_1 = f_\mu(0.5)$ and $q_2 = f_\mu(1)$. Therefore, we can move across the DK parameter space $(q_1, q_2)$ varying $\mu$, going from the inactive phase ($\mu < \mu_c^{\infty}$) to the active one ($\mu > \mu_c^{\infty}$), and no bistable phase is possible. We find that the transition happens at a critical $\mu_c^\infty = 0.85(7)$. Note that, in the active phase, a completely empty state ($p_1 = 0$) remains trivially empty at all times. This behavior is however unstable, because any $p_1 > 0$ leads to percolation (i.e., $p_{1,c} = 0$), and we therefore do not classify the active phase as bistable. At criticality, and for $p_1 \ll 1$, the density grows as $n \sim t^\theta$ with $\theta = 0.3(0)$, as expected for the DP universality class \cite{hinrichsen2000non}. See Supplementary Fig.~2 for details.
	
	In the infinite-range limit $\alpha \to 0$, and more generally for $\alpha \le 1$, the factor $\mathcal{N}_{\alpha, L}$ in Eq.~\eqref{eq. loc density} diverges as $L \to \infty$. Correspondingly, spatial stochastic fluctuations are suppressed, that is, all sites $i$ share the same occupation probability $p_{i,t+1} = p_t$ and density $n_{i,t} = n(t) = p_t$. Therefore, in this limit the dynamics reduces to the deterministic $0$-dimensional recurrence relation 
	\begin{equation}
	n(t+1) = f_\mu \left[ n(t) \right].
	\label{eq. FP1}
	\end{equation}
	The system asymptotic behaviour can then be understood from the analysis of the fixed points (FPs) of the equation $x = f_\mu(x)$, which is detailed in Section Methods.
	
	\begin{figure*}[t]
		\begin{center}
			\includegraphics[width=\linewidth]{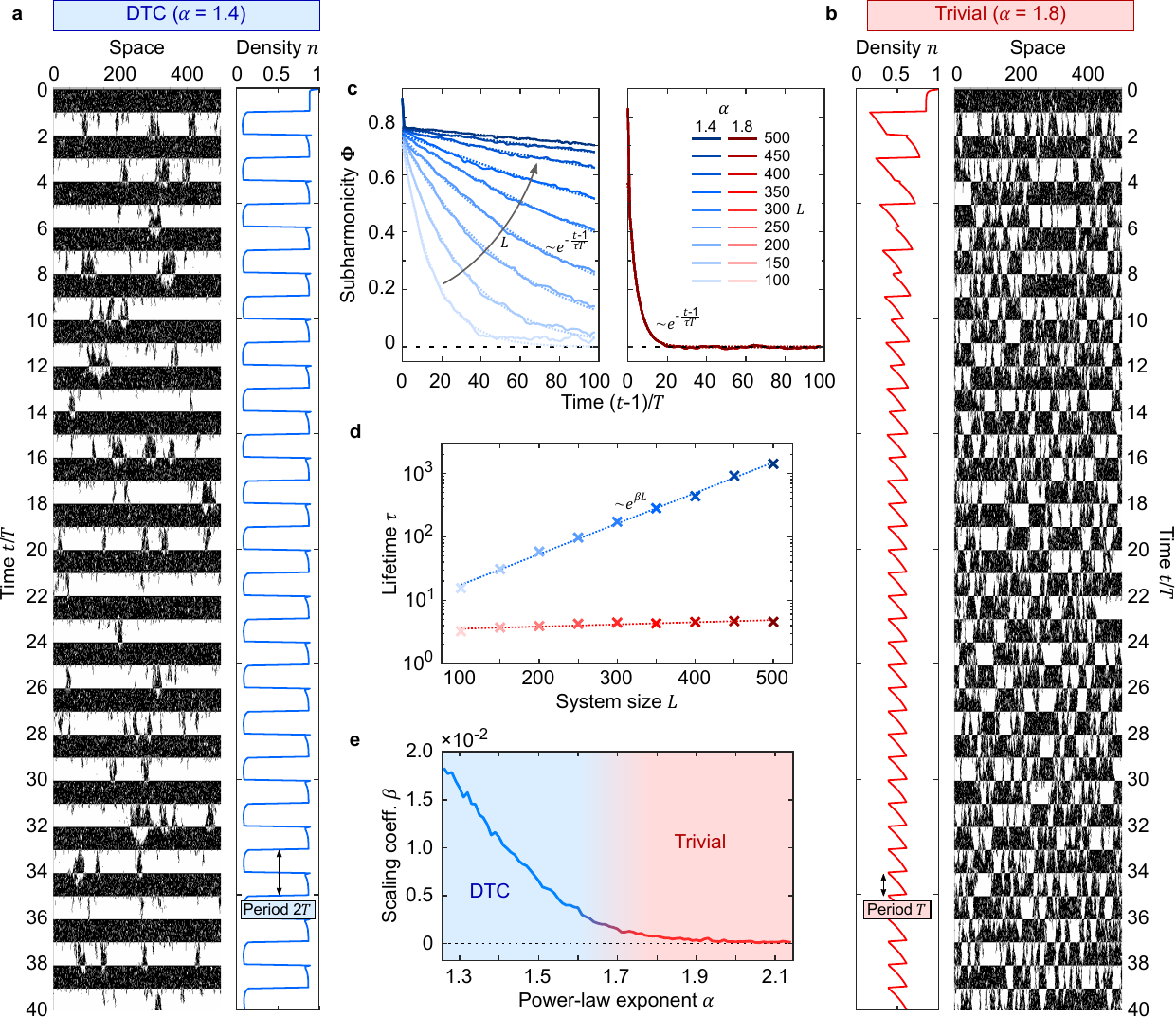}\\
		\end{center}
		\vskip -0.5cm \protect
		\caption{
			\textbf{Discrete time crystals in Floquet long-ranged directed percolation.}
			(a,b) Single instances of the Floquet DP, alongside with the density $n$ averaged over multiple independent runs, for $L = 500$ sites. (a) For a power-law exponent $\alpha = 1.4$, $n$ oscillates subharmonically with a period that is twice that of the drive, whereas, for $\alpha = 1.8$, $n$ eventually picks the periodicity $T$ enforced by the drive.
			(c) For finite system sizes $L$, the subharmonicity $\Phi(t)$ decays as $\Phi(t) \sim \exp(-\frac{t-1}{\tau T})$ due to the accumulation of phase slips, and, after a few time scales $\tau T$, the density $n$ synchronises with the drive and oscillates with period $T$. Exponential fits (dotted lines) can be used to extrapolate the lifetime $\tau$ of the subharmonic response, on which a scaling analysis is performed in (d). For $\alpha = 1.4$ (blue), the lifetime $\tau$ scales exponentially with the system size, $\tau \sim e^{\beta L}$, whereas no such a scaling is found for $\alpha = 1.8$. The scaling coefficient is again found from an exponential fit (dotted line), and plotted in (e) versus the power-law exponent $\alpha$. For small $\alpha$, that is long-ranged enough DP, the scaling coefficient $\beta$ is finite, indicating that in the thermodynamic limit $L \to \infty$ the subharmonic response is persistent and a DTC with infinite autocorrelation time emerges. On the contrary, $\beta \approx 0$ for large $\alpha$, indicating a trivial dynamical phase in which no stable subharmonic dynamics is established. Here, we considered $p_1 = 1, \mu = 0.9, p_d = 0.02, T = 20$ and $R = 2000$.}
		\label{fig3}
	\end{figure*}
	
	\textbf{Floquet percolation and time crystals.}
	We have established that long-range correlated local densities $\{n_{i,t}\}_i$ give rise to a bistable phase. We now show how, in a Floquet DP with periodically modulated update rules, this phase intimately relates to the emergence of a classical DTC. In this phase, as we shall see, the density $n$ displays oscillations over a period larger than that of the drive and up to a time that, thanks to the long-range interactions and despite the presence of multiple sources of noise, is exponentially large in the system size, a feature that would generally be forbidden in short-ranged PCA \cite{yao2020classical}. In the thermodynamic limit $L \to \infty$, these subharmonic oscillations are therefore persistent, that is, the system autocorrelation time diverges to infinity, breaking the time-translational symmetry and proving a classical DTC in a Floquet PCA.
	
	In the spirit of keeping the model as simple as possible, we consider a minimal Floquet drive in which, after every $T$ iterations of the DP in Eqs.~\eqref{eq. loc density} and \eqref{eq. f}, empty sites are turned into occupied ones and vice versa, making the full equations of motion periodic with period $T$. As a further source of imperfections, adding to the underlying noisy DP, we also account for faulty swaps with probability $p_d$. More explicitly, the Floquet drive consists of the following transformation
	\begin{equation}
	s_{i, 1+kT} \to
	\begin{cases}
	1-s_{i, 1+kT} & \text{with probability} \ 1-p_d \\
	s_{i, 1+kT} & \text{with probability} \ p_d.
	\end{cases}
	\label{eq. Floquet}
	\end{equation}
		
	In Fig.~\ref{fig3}(a,b) we show the spatio-temporal pattern of single instances of the Floquet DP, alongside with the density $n$ averaged over several independent runs. If the DP is short-ranged enough, the spatio-temporal pattern at long times looks similar from one Floquet period to the next, that is the density $n$ synchronises with the drive and eventually picks a periodicity $T$. On the contrary, for a long-ranged enough DP, the system keeps alternating at every period between a densely occupied regime and a sparsely occupied one, and $n$ oscillates with period $2T$, that is, the system breaks the discrete time-translation symmetry of the equations of motion.
	 
	When using the tag `classical DTC', special care should be reserved for showing the defining features of this phase, namely its rigidity and persistence~\cite{yao2020classical}. Our system is rigid in the sense that it does not rely on fine-tuned model parameters, e.g., $\mu$, $\alpha$ or the initial density $p_1$, and that noise, either in the form of the inherently stochastic underlying DP or of a small but non-zero Floquet defect density $p_d$, does not qualitatively change the results. Moreover, in the limit $L \to \infty$, our DTC is truly persistent. Indeed, one might expect that the accumulation of stochastic mistakes introduces phase slips and eventually leads to the (possibly slow but unavoidable) destruction of the subharmonic response. Although this expectation is generally correct for short-ranged DP models, including our model at large $\alpha$, it can fail for long-ranged DP models.
	
	To show that, in the limit $L\to \infty$, the lifetime of our DTC is infinite, we perform a scaling analysis comparing results for increasing system sizes $L$. First, we introduce an order parameter $\Phi(t)$, henceforth called \emph{subharmonicity}, that is defined at stroboscopic times $t = 1, 1+T, 1+2T, \dots$ as
	\begin{equation}
	\Phi(t) = (-1)^{\frac{t-1}{T}} \left[n(t) - n(t+T)\right].
	\end{equation}
	If the density $n$ oscillates with the same period $T$ as the drive, then $n(t) = n(t+T)$ and $\Phi(t) = 0$. On the contrary, if $n$ oscillates with a doubled period $2T$, then $n(t=1+kT)$ is positive and negative for even and odd $k$, respectively, and $\Phi(t)$ is finite and maintains a constant sign. Therefore, $\Phi(t)$ is a suitable diagnostics to track the degree of subharmonicity of $n$ in time, and to perform the scaling analysis.
	
	In Fig.~\ref{fig3}(c) we show $\Phi(t)$ for various system sizes $L$. For both $\alpha = 1.4$ and $\alpha = 1.8$, the subharmonicity decays exponentially in time, $\Phi(t) \sim \exp(-\frac{t-1}{\tau T})$. As shown in Fig.~\ref{fig3}(d), these two values of $\alpha$ are however crucially different in how the lifetime $\tau T$ scales with the system size. In fact, $\tau T$ is approximately independent of $L$ for $\alpha = 1.8$, whereas it scales exponentially as $ \tau \sim \exp(\beta L)$ for $\alpha = 1.4$, for which the decay of the subharmonicity is therefore just a finite-size effect. The scaling coefficient $\beta$ quantifies the time crystallinity of the system, and can thus be used to obtain a full phase diagram as a function of the power-law exponent $\alpha$, in Fig.~\ref{fig3}(e). We observe a phase transition between a DTC and a trivial phase at $\alpha \approx 1.7$. That is, if the DP is sufficiently long-ranged ($\alpha \lessapprox 1.7$), $\beta$ is finite and in the thermodynamic limit $L \to \infty$ the subharmonic response extends up to infinite time, as required for a true DTC. In contrast, for a shorter-range DP ($\alpha \gtrapprox 1.7$), $\beta \approx 0$ independently of $L$ and the subharmonic response is always dynamically destroyed.
	
	\begin{figure*}[t]
		\begin{center}
			\includegraphics[width=\linewidth]{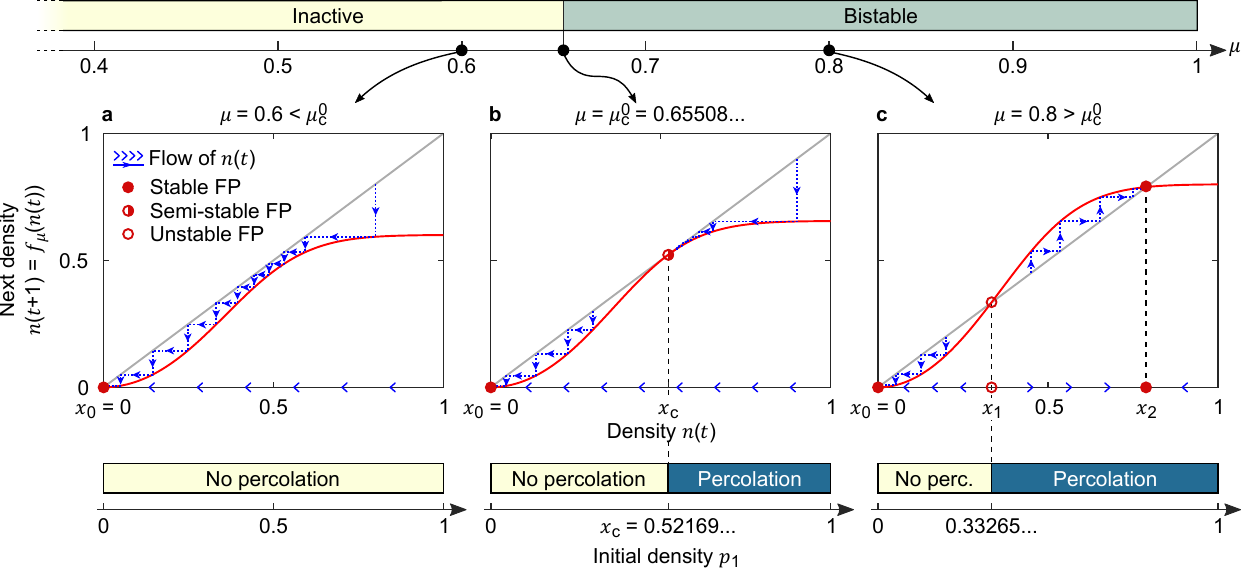}\\
		\end{center}
		\vskip -0.5cm \protect
		\caption{
			\textbf{Graphical fixed-point analysis.} For a power-law exponent $\alpha < 1$, the dynamics of Eq.~\eqref{eq. FP1} is understood from the FP analysis of the equation $x = f_\mu(x)$. (a) For a control parameter $\mu < \mu_c^0 = 0.6550(8)$, the system is inactive, corresponding to a single FP $x_{0} = 0$: at long times, the system ends up in the empty, absorbing state with state variables $s_{i} = 0$ for all sites $i$. (b) At the critical point $\mu = \mu_c^0$, a new semi-stable FP emerges at $x_c = 0.5216(9)$, that is unstable from his left and stable on his right. (c) Increasing $\mu$ above $\mu_c^0$, the semi-stable FP splits into an unstable FP $x_1 < x_c$ and a stable FP $x_2 > x_c$. Depending on whether the initial density $p_1$ is $< x_1$ or $> x_1$, the system flows towards density $n = x_0 = 0$ or $n = x_2 > 0$, respectively, indicating bistability.}
		\label{figM1}
	\end{figure*}

	\section{Discussion}
	We have shown that long-range DP and its Floquet variant can give rise to a bistable phase and a DTC, respectively.
	At the core of our model in Eq.~\eqref{eq. loc density} and Eq.~\eqref{eq. f} is the idea that the occupation of a given site depends on the state of \emph{all} the other sites at the previous time. In this sense, our model is reminiscent of some SIR type models of epidemic spreading in which not only a sick site can infect a susceptible site, but several infected sites can also cooperate to weaken a susceptible site, and finally infect it \cite{bizhani2012discontinuous, janssen2004generalized}. This cooperation mechanism among an infinite number of parent sites, rather than a finite one as considered in previous works on long-ranged DP \cite{janssen1999levy, hinrichsen1999model}, is the key feature allowing the emergence of the bistable phase, that finds a transparent explanation in the infinite-range limit $\alpha \to 0$, where it corresponds to the equation $x = f_\mu(x)$ having two stable FPs. Bistability also provides intuition on the origin of the DTC, to which it is deeply connected. Indeed, the Floquet drive in Eq.~\ref{eq. Floquet} switches the system from a densely occupied regime to a sparsely occupied one (and vice versa). If the underlying DP is bistable, these regimes fall each within different basins of attraction, and can therefore be both stabilized by the contractive dynamics \cite{gambetta2019discrete, Gambetta2019}. Ultimately, this double stabilization facilitates the establishment of the DTC with infinite autocorrelation time. Remarkably, this mechanism does not rely on the equations of motion being perfectly periodic, as required for DTCs in closed MBL systems \cite{von2016absolute}, and we expect that infinite autocorrelation times could be maintained even in the presence of aperiodic variations of the drive (although the nomenclature should be revised in this case, since the underlying discrete time symmetry would only be present on average but not for individual realizations). This is in contrast to DTCs in closed MBL systems \cite{von2016absolute}, in which the non-ergodic dynamics hinges on the peculiar mathematical structure of the Floquet operator which, in turns, relies on the underlying equations being perfectly periodic.
	
	The intimate connection between bistability and DTC is however not a strict duality, and the boundaries of the two phases, in the equilibrium and non-equilibrium phase diagrams, respectively, do not coincide. For instance, in our analysis we found that for $\mu = 0.9$ the bistable phase extends up to $\alpha \approx 1.6$, whereas the DTC stretches slightly farther, up to $\alpha \approx 1.7$. The origins of this imperfect correspondence can be traced back to two competing effects.
	On the one hand, bistability may not be sufficient to stabilize a DTC. This can already be understood in the limit $\alpha \to 0$, in which the asymmetry of $f_\mu$ and of its FPs does not guarantee the Floquet driving to switch the density $n$ from one basin of attraction to the other, that is,  across the critical probability $p_{1,c}$. This issue becomes even more relevant for larger $\alpha$, for which the asymmetry is possibly accentuated and $p_{1,c}$ can approach $0$ (see for instance Supplementary Fig.~1). On the other hand, a perfect bistability may not even be necessary for a DTC to exist. In fact, for the stabilization of a DTC, it may be sufficient that, of the densely and sparsely occupied regimes of the underlying DP, only one is stable, and the other is just weakly unstable (that is, metastable), meaning that the timescales of the dynamics of the density $n$ in the two regimes are very different. Loosely speaking, the stability of one regime might be able to compensate for the weaker instability of the other, resulting in an overall stable DTC. The asymmetry of the underlying DP and the mismatch between the bistable phase and the DTC highlight the purely dynamical nature of the latter, that cannot 'piggy-back` on any underlying symmetry.
	
	While these considerations are model and parameters dependent, and it is ultimately up to numerics to find the bistable and the DTC phases, what is universal and far reaching here is the concept that long-ranged DP, and PCA more generally, can host novel dynamical phases such as DTCs. As Yao and collaborators recently pointed out \cite{yao2020classical}, long autocorrelation times are in fact generally unexpected in $1+1$-dimensional PCA, because imperfections and phase slips can nucleate, spread and destroy the order. Our work proves that this fate can be avoided, and time-crystalline order established, in long-ranged PCA. These systems enable in fact an error correction mechanism, in our case intimately related to the bistability, that would be impossible if correlations were limited to a finite radius. We may speculate that, in the physical picture of a Hamiltonian system coupled to a bath, this defect suppression would correspond to the cooling rate being larger than the heating rate.
	
	In conclusion, we have studied the effects of long-range correlated update rules in a model of DP, which we built from an analogy with the prototypical (but short-ranged) DK PCA. First, we proved that, beyond the standard active and inactive phases, a new bistable phase emerges in which the system at long times is either empty or finitely occupied depending on whether it was initially sparsely or densely occupied. Second, in a Floquet DP with periodic modulation of the update rules, we showed that this bistable phase intimately connects with a DTC phase, in which the density oscillates with a period twice that of the drive. In this DTC phase, the autocorrelation time scales exponentially with the system size, and in the thermodynamic limit a robust and persistent breaking of the discrete time-translation symmetry is established.
	
	As an outlook for future research, further work on the Floquet DP should better assess the nature of the transition between the DTC and the trivial phase, characterise more systematically the phase diagram in other directions of the parameter space, and, most interestingly, address the role of dimensionality. Indeed, it is well-known that dimensionality can facilitate the establishment of ordered phases of matter at equilibrium, and the question whether this is the case also out-of-equilibrium remains open. A positive answer to this question is suggested by the fact that, in $D+1$-dimension with $D \ge 2$, bistability can emerge even in short-ranged models of DP \cite{janssen2004generalized, lubeck2006tricritical, grassberger2006tricritical}. Another interesting question regards the fate of chaos and damage spreading in long-ranged DP ~\cite{Martins1991evidence}. Further research should then aim to gain analytical intuition into the problem. For instance, the critical $\alpha$ separating the various phases may be located using a field theoretical approach, which has been successful in similar contexts in the past \cite{hinrichsen1999model}. Finally, on a broader perspective, our work paves the way towards the study of non-equilibrium phases of matter in the uncharted territory of Floquet PCA, with a potentially very broad range of applications throughout different branches of science. As a timely example, Floquet PCA may provide new insights into the understanding of seasonal epidemic spreading and periodic intervention efficacy.

	\section{Methods}
	Here, we provide further technical details on our work. 
	In Eq.~\eqref{eq. loc density}, we considered as distance $r_{i,j}$ between sites $i$ and $j$
	\begin{equation}
	r_{i,j} = \frac{L}{\pi} \left|\tan(\pi\frac{i-j}{L})\right|,
	\label{eq. r_ij}
	\end{equation}
	where the tangent accounts for periodic boundary conditions and makes the distance of the fartherst sites with $|i-j| = L/2$ artificially diverge. This divergence is expected to reduce finite-size effects without changing the underlying physics, that is in fact dominated by sites with $|i-j| \ll L$, for which we get a natural $r_{i,j} \approx |i-j|$. Indeed, as we checked, similar results are obtained with $r_{i,j} = \min (|i-j|, L-|i-j|)$. The Kac-like normalization factor $\mathcal{N}_{\alpha, L}$ reads instead	
	\begin{equation}
		\mathcal{N}_{\alpha, L} = \sum_{j = 1}^{L} \left(r_{\frac{1}{2},j}\right)^{-\alpha}.
	\end{equation}
	
	The phenomenology of the bistable phase can be understood from a graphical FP analysis of the equation $f_\mu(x) = x$ illustrated in Fig.~\ref{figM1}, which explains the dynamics for $\alpha < 1$. Three scenarios are possible, and interpreted in terms of the ways the graph of the function $f_\mu$ intersects with the bisect. (i) Inactive -- if $\mu < \mu_c^{0}$, the only FP is $x_0 = 0$, which is stable and corresponds to a completely empty state. The system moves towards this FP and $p_t \xrightarrow{t \to \infty} 0$. (ii) Critical -- if $\mu = \mu_c^0$, a new semi-stable FP emerges at $x_c$, which is attractive from its right and repulsive on its left. (iii) Bistable -- if $\mu > \mu_c^0$, the semi-stable FP splits into an unstable FP $x_1 > x_0$ and a stable FP $x_2 > x_1$. In this case, the system will reach either the unoccupied FP $x_0 = 0$ or the finitely occupied FP $x_2 > 0$ depending whether $p_1 < x_1$ or $p_1 > x_1$, respectively. That is, the system is bistable, and the critical initial probability separating its two basins of attraction is $p_{1,c} = x_1$ (see also Supplementary Fig.~1). The critical value $\mu_c^0$ is obtained numerically solving for the condition of tangency between the graph of $f_\mu$ and the bisect, and gives $\mu_c^0 = 0.6550(8)$ and $x_c = 0.5216(9)$. For $\mu > \mu_c^0$, the FPs $x_1$ and $x_2$ are found solving for $f_\mu(x) = x$, and, for instance, we find we find $x_1 = 0.3326(5)$ and $x_2 = 0.7890(9)$ for $\mu = 0.8$.
	
	The FP analysis also clarifies the general features of $f_\mu$ that allow for the emergence of bistability, that is in fact not contingent on the choice of $f_\mu$ made in Eq.~\eqref{eq. f}. Indeed, the only requirement is that, for some parameter(s) $\mu$, the equation $f_\mu(x) = x$ has three FPs $x_0 < x_1 < x_2$, of which $x_0$ and $x_2$ are stable, whereas $x_1$ is unstable. Put simply, $f_\mu$ should be a nonlinear function with a graph looking qualitatively as that of Fig.~\ref{figM1}(c). This condition guarantees a bistable phase for $\alpha < 1$, which can then possibly extend to $\alpha \ge 1$ and, in the presence of a Floquet drive, facilitate the establishment of a DTC.
	
	Finally, note that higher resolution and smaller fluctuations could be achieved in the figures throughout the paper if simulating larger system sizes $L$ and/or considering a larger number of independent runs $R$. This could, for instance, allow a more accurate characterisation of both the equilibrium and the non-equilibrium phase diagrams of our model, which could be explored in other directions of the parameter space for varying $\alpha, \mu$, $p_d$ and $T$. This would, however, require a formidable numerical effort, and goes therefore beyond the scope of this work. As a reference, for instance, the generation of Fig.~\ref{fig3}(e) for the parameters considered therein requires a computing time of approximately $4 \times 10^3$ hours per $3$ GHz core.
	
	\textbf{Data availability.}
	No datasets were generated or analysed during the current study.
	
	\textbf{Acknowledgements.}
	We are very thankful to P.~Grassberger for insightful comments on the manuscript. J.~K.~thanks Kim Christensen for introducing him to the theory of percolation. We acknowledge support from the Imperial-TUM flagship partnership. A.~P.~acknowledges support from the Royal Society. A.~N.~holds a University Research Fellowship from the Royal Society and acknowledges additional support from the Winton Programme for the Physics of Sustainability.
	
	\textbf{Author contributions.}
	J.~K.~initiated the project suggesting to investigate DTCs in long-ranged DP models and to take inspiration from the DK model. A.~P.~proposed the model and performed the computations. A.~N.~made critical contributions to the analysis of the results and the preparation of the manuscript.
	
	\textbf{Code availability.}
	The codes that support the findings of this study are available at https://figshare.com/articles/software/Code/13468836.

	%

	\clearpage
	
	\setcounter{equation}{0}
	\setcounter{figure}{0}
	\setcounter{page}{1}
	\thispagestyle{empty} 
	\makeatletter 
	\renewcommand{\figurename}{Supplementary Fig.}
	\renewcommand{\thefigure}{\arabic{figure}}
	\renewcommand{\theequation}{S\arabic{equation}}
	\setlength\parindent{10pt}
	
	\onecolumngrid
	
	\begin{center}
		{\fontsize{12}{12}\selectfont
			\textbf{Supplementary Information for\\``Bistability and time crystals in long-ranged directed percolation"\\[5mm]}}
		{\normalsize Andrea Pizzi, Andreas Nunnenkamp, and Johannes Knolle \\[1mm]}
	\end{center}
	\normalsize
	
	\clearpage
	
	\subsection*{Supplementary Note 1: Role of the initial density.}
	First, we investigate our system in yet another direction in the parameter space, that is that of the initial occupation probability (or density) $p_1$. In the main text, we have in fact seen that the system behaviour in the bistable phase drastically depends on whether $p_1$ is `small' or `large', and, having so far limited our examples to $p_1 = 0.01$ and $p_1 = 1$, we now better assess what do `small' and `large' mean. In Supplementary Fig.~\ref{figS1} we consider the entire range of $p_1$ from $0$ to $1$, and look for the critical density $p_{1,c}$ separating the two basins of attraction: $n$ does and does not decay to $0$ for $p_1 < p_{1,c}$ and $p_1 > p_{1,c}$, respectively. The critical probability at $\alpha \to 0$ coincides with the unstable fixed point $x_1$ of the equation $x = f_\mu(x)$, as explained in the main text, and is reported in the plots as a reference. For $\mu = 0.8$ (left), we find that $p_{1,c} \approx x_1$ for all the $\alpha \lessapprox 1.5$, whereas for larger $\alpha$ the system enters the inactive phase. For $\mu = 0.9$, we observe that $p_{1,c}$ changes smoothly with $\alpha$ from $x_1$ at $\alpha = 1$ to $0$ at $\alpha \approx 1.6$, when the system enters the active phase, see also Fig.~2(d,e).
	
	As we have show here, the critical probability $p_{1,c}$ of the bistable phase generally falls in the bulk of the range $(0,1)$, ultimately because the FP $x_1$ does. This is an important feature of our model, because it means that both the possible behaviours of the bistable phase, that is percolating and not, have a broad range of $p_1$ in which they are stable. On the one hand, this means that the results of Fig.~1 in the main text are not contingent on the choice of $p_1 = 0.01$ and $1$, but would rather be analogue for other choices of $p_1 < p_{1,c}$ and $p_1 > p_{1,c}$. On the other hand, once the Floquet drive is included, such a broad stability region enables the DTC robustness to noise.
	
	\begin{figure}[h]
		\begin{center}
			\includegraphics[width=0.5\linewidth]{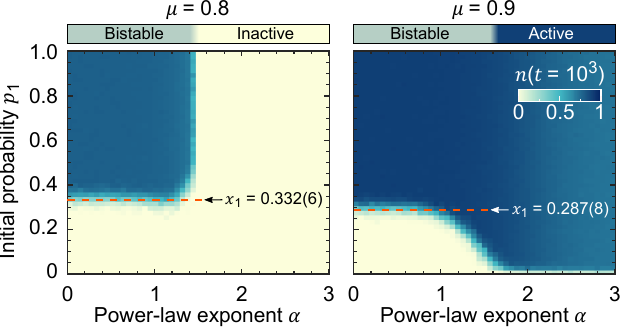}\\
		\end{center}
		\vskip -0.5cm \protect
		\caption{
			\textbf{Role of the initial density in the bistable phase}. Long-time density $n(t = 10^3)$ in the plane of the power-law exponent $\alpha$ and the initial occupation density $p_1$, for a value of the control parameter $\mu > \mu_c^0 = 0.6550(8)$. For sufficiently small $\alpha$, the system is in the bistable phase, meaning that $n(10^3)$ can be either finite or not depending on $p_1$, the two behaviours being separated by a critical probability $p_{1,c}$. For $\alpha \to 0$, the critical probability $p_{1,c}$ corresponds to $x_1$, the unstable FP of the equation $x = f_\mu(x)$, which is reported as a reference. For $\mu = 0.8$ (left), we observe that $p_{1,c} \approx x_1$ for $\alpha$ up to $\approx 1.5$, above which the system enters the inactive phase. For $\mu = 0.9$ (right), instead, the critical $p_{1,c}$ decreases smoothly with $\alpha$, reaching $0$ at $\alpha \approx 1.6$, at which the system enters the active phase. Here, $L = 500$ and $R = 100$.}
		\label{figS1}
	\end{figure}
	
	\subsection*{Supplementary Note 2: Short-range DK limit.}
	Second, we investigate in more detail the limit $\alpha \to \infty$ already treated in the main text. In such a limit, our model of DP recasts into the DK model, upon replacing $q_1 = f_\mu(0.5) = \mu \tanh(1)$ and $q_2 = f_\mu(1) = \mu \tanh(4)$. Varying $\mu$, one can therefore move across the DK parameter space $(q_1, q_2)$ along the line $q_2 = \frac{\tanh 4}{\tanh 1} q_1$, and therefore across the phase boundary between the active and inactive phases. In Supplementary Fig.~\ref{figS2}, we plot the time series of the density $n$ for various values of the control parameter $\mu$. Reiterating from the main text, what we find is that for $\mu = \mu_c^\infty = 0.85(7)$ the density grows as a power law $\sim t^\theta$ with $\theta = 0.3(0)$ as expected in the DP universality class. For $\mu > \mu_c^\infty$ the density $n$ grows in time, whereas for $\mu < \mu_c^\infty$ it rather decays to $0$, signalling the active and inactive phases, respectively.
	
	Note that the setting in which DP is studied is usually that of an initial condition with one `seed site' being occupied, and an infinitely large system size $L$, corresponding to a density $1/L \to 0$. In our case, the choice of the initial condition with occupation probabilities $p_1$ corresponds to a possibly small but still finite initial density $p_1$, so that in the limit $L \to \infty$ there will always be infinitely many seed sites. In the active phase, the clusters originating from many of these sites will grow and expand in time, eventually merging together and leading to a saturation of $n$ at long-times. In particular, the universal power-law growth at criticality can be observed only for $p_1 \ll 1$, and it extends for a finite time, before saturation eventually sets in.
	
	\begin{figure}[h]
		\begin{center}
			\includegraphics[width=0.5\linewidth]{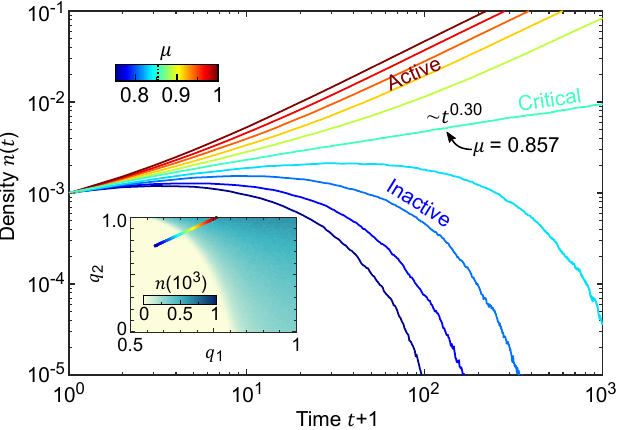}\\
		\end{center}
		\vskip -0.5cm \protect
		\caption{
			\textbf{Limit of short-range directed percolation}. In the short-range limit $\alpha \to \infty$, our model of DP maps to the DK model. In this limit, we show the dynamics of the density $n$ for various values of the control parameter $\mu$. For $\mu < \mu_c^\infty$ ($\mu > \mu_c^\infty$), the system does not (does) percolate, that is $n$ does (does not) decay to $0$ at long-times. At the critical point $\mu = \mu_c^\infty \approx 0.857$ (indicated with a dotted line in the colorbar), $n$ grows as a power-law $\sim t^\theta$, with $\theta \approx 0.31$ as expected in the DP universality class. Analogue results are obtained for other choices of the initial density $p_1$, since in the short-ranged DK limit no bistable phase exists. Here, we consider the following parameters $L = 10^3$, $p_{1} = 10^{-3}$ and $R = 10^4$. In the inset, we report for reference Fig.~1(d), highlighting the line spanned in the $(q_1, q_2)$ DK parameter space when varying $\mu$.}
		\label{figS2}
	\end{figure}
	
\end{document}